\documentclass[twocolumn, iop, apj, twocolappendix, numberedappendix, appendixfloats]{emulateapj}
\usepackage{CJKutf8}
\usepackage{threeparttable}
\usepackage{multirow}
\usepackage{times}
\usepackage{enumitem}
\usepackage{url} 
\usepackage{color}
\usepackage{amsmath,bm}
\usepackage{subfigure}
\usepackage{natbib}
\usepackage{graphicx}
\usepackage{graphics}
\usepackage{hyperref}                             
\usepackage{amsmath}
\usepackage{bm}
\usepackage{subfigure}
\usepackage{array}
\usepackage{booktabs}
\usepackage{lineno}

\shorttitle{The tilt of the velocity ellipsoid}
\shortauthors{Sun et al.}
\submitted{Submitted to ApJ;\,\,\,\,Accepted June 02, 2023}
\begin{document}

\title{The tilt of the velocity ellipsoid of different Galactic disk populations}
\author{ Weixiang Sun\textsuperscript{1,2}}
\author{ Han Shen\textsuperscript{1}}
\author{ Xiaowei Liu\textsuperscript{1,2}}

\altaffiltext{1}{South-Western Institute for Astronomy Research, Yunnan University, Kunming 650500, People's Republic of China; {\it w.sun@mail.ynu.edu.cn {\rm (WXS)}; x.liu@ynu.edu.cn (XWL)}}
\altaffiltext{2}{Corresponding authors}

\begin{abstract}

The tilt of the velocity ellipsoid is a helpful tracer of the gravitational potential of the Milky Way.
In this paper, we use nearly 140,000 RC stars selected from the LAMOST and {\it Gaia} to make a detailed analysis of the tilt of the velocity ellipsoid for various populations, as defined by the stellar ages and chemical information, within 4.5\,$\leq$\,$R$\,$\leq$\,15.0\,kpc and $|Z|$\,$\leq$\,3.0\,kpc.
The tilt angles of the velocity ellipsoids of the RC sample stars are accurately described as $\alpha$ = $\alpha_{0}$\,$\mathrm{arctan}$\,($Z$/$R$) with $\alpha_{0}$ = (0.68 $\pm$ 0.05).
This indicates the alignment of velocity ellipsoids is between cylindrical and spherical, implying that any deviation from the spherical alignment of the velocity ellipsoids may be caused by the gravitational potential of the baryonic disk.
The results of various populations suggest that the $\alpha_{0}$ displays an age and population dependence, with the thin and thick disks respectively values $\alpha_{0}$ = (0.72 $\pm$ 0.08) and $\alpha_{0}$ = (0.64 $\pm$ 0.07), and the $\alpha_{0}$ displays a decreasing trend with age (and [$\alpha$/Fe]) increases, meaning that the velocity ellipsoids of the kinematically relaxed stars are mainly dominated by the gravitational potential of the baryonic disk.
We determine the $\alpha_{0}$ -- $R$ for various populations, finding that the $\alpha_{0}$ displays oscillations with $R$ for all the different populations.
The oscillations in $\alpha_{0}$ appear in both kinematically hot and cold populations, indicating that resonances with the Galactic bar are the most likely origin for these oscillations.

\end{abstract}


\section{Introduction}

The distribution of the mass of the Galaxy is a powerful tool for constraining the formation and evolution histories of the Milky Way \citep[e.g.,][]{Binney2008, Smith2009, Everall2019}.
However, the mass that can emit detectable electromagnetic radiation only accounts for a small portion of the total mass of the Milky Way \citep[e.g.,][]{Huang2016, Eilers2019, Ablimit2020}, owing to which astronomers tend to use indirect methods to measure it.
One of the most powerful methods is to measure the ellipsoidal distribution of stellar velocity \citep[e.g.,][]{Eddington1915, Smith2009, Hagen2019}.

The velocity ellipsoids are widely used to detect the global gravitational potential of the Milky Way \citep[e.g.,][]{Budenbender2015, Hagen2019, Everall2019}.
Those results indicate that the tilt angle of the velocity ellipsoid can be described as $\alpha$ = $\alpha_{0}$\,$\mathrm{arctan}$\,($Z$/$R$) \citep[e.g.,][]{Binney2014, Budenbender2015}, in which $\alpha_{0}$\,=\,1.0 means exact spherical alignment, while $\alpha_{0}$\,$<$\,1.0 implies the velocity ellipsoids are tilted towards cylindrical alignment.
The tilted towards cylindrical alignment is likely related to the effect of the gravitational potential of the baryonic disk.\citep[e.g.,][]{Hagen2019, Everall2019}.

Binney et al. ({\color{blue}{2014}}) reported that $\alpha_{0}$ $\sim$ 0.8 for the local RAVE sample \citep{Steinmetz2006}.
However, the results from the SEGUE G-dwarf sample tend to suggest a more spherically aligned result, with $\alpha_{0}$\,$\sim$\,0.9 \citep{Budenbender2015}.
In recent studies, Hagen et al. ({\color{blue}{2019}}) found that the velocity ellipsoids trend to cylindrically aligned outer the Solar radius, but changes spherically aligned in the inner disk, revealing the Galactic space dependence of the velocity ellipsoids.
A similar result is further reported by Gaia Radial Velocity Spectrometer (RVS) samples \citep{Everall2019}, it suggests that the velocity ellipsoids show Galactic altitude dependence, with higher Galactic altitude inclined to a more spherically aligned.
In addition, the results of different populations indicate that the distributions of the velocity ellipsoids are distinguished for different populations, with the thin disk population likely tend to deviate obviously from spherical alignment \citep[e.g.,][]{Binney2014, Hagen2019, Everall2019}.
However, a detailed analysis of the velocity ellipsoid based on different stellar ages and abundance populations is still not well characterized since the sample used in previous studies did not provide accurate measurements of stellar age and metallicity \citep[e.g.,][]{Hagen2019, Everall2019}.
At present, a large sample of the red clump (RC) stars \citep{Huang2020} from LAMOST and Gaia surveys presents a compelling opportunity for conducting research in the field.
Leveraging this sample, it is possible to perform an exhaustive analysis of the velocity ellipsoids characterizing the Galactic disk, thereby enhancing our comprehension of the gravitational potential distributions governing the Milky Way.

This paper is organized as follows: In Section\,2, we describe the data used in this paper.
The calculation of the tile angle of the velocity ellipsoid is discussed in Section\,3, and present our results in Sections 4 and 5.
Finally, we summarized our main conclusions in Section\,6.

\begin{figure}[t]
\begin{center}
\includegraphics[width=8.9cm]{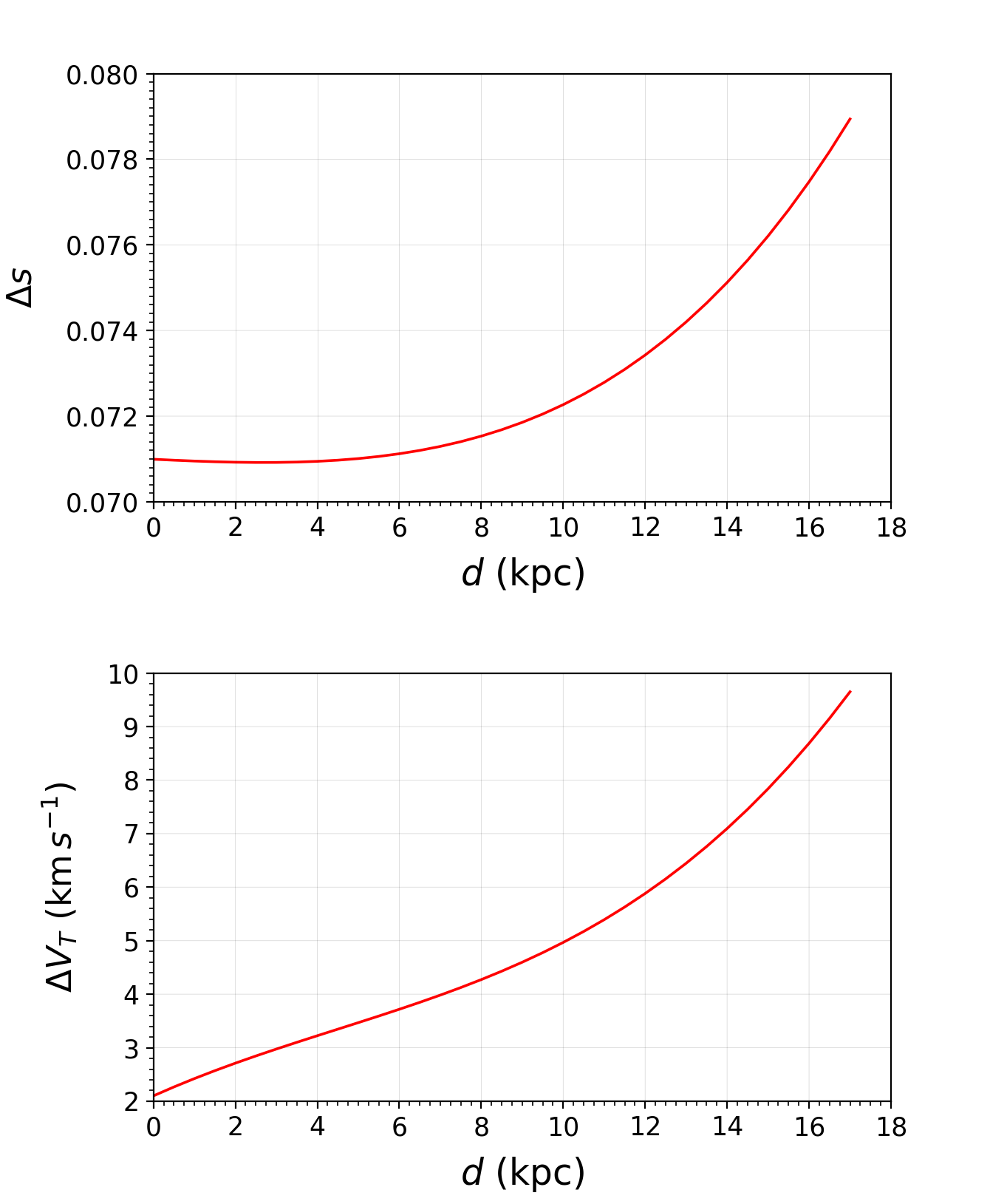}
\caption{The relative distance error ($\Delta s$, is defined as the ratios of the distance errors to distances) as a function of distance (upper panel), and the tangential velocity error ($\Delta V_{T}$, is defined as the dispersion of the 1000 times Markov Chain Monte Carlo (MCMC) sampling of $V_{T}$) as a function of distance (bottom panel).
The red line in each panel is the best fit for all the RC sample stars.}
\end{center}
\end{figure}

\section{Data}

In this work, we mainly use a sample with 137,448 red clump (RC) stars \citep[e.g.,][]{Huang2020} from LAMOST survey \citep[e.g.,][]{Cui2012, Deng2012, Liu2014, Yuan2015}, with the uncertainties, of the effective temperature ($T_{\rm eff}$), line-of-sight velocity ($V_{\rm r}$), surface gravity (log\,$g$), [Fe/H] and [$\alpha$/Fe], are typically around, 100\,K, 5\,km\,s$^{-1}$, 0.10\,dex, 0.10$-$0.15\,dex and 0.03$-$0.05\,dex \citep[e.g.,][]{Xiang2017, Huang2018a}.
Due to the standard-candle nature of the RC stars, the distance accuracy is better than around 5\%--10\% (see the upper panel of Fig.\,1).
To further improve the accuracy of the kinematic estimation, we update the astrometric parameters of the sample stars to Gaia DR3 \citep[e.g.,][]{Gaia2022a, Gaia2022b, Recio2022}.
Therefore, the uncertainty of the measured tangential velocity ($V_{T}$ = $d$\,$\sqrt{\mu_{\alpha}^{2} + \mu_{\delta}^{2}}$, where $d$ is the distance from the Sun, $\mu_{\alpha}$ and $\mu_{\delta}$ are respectively proper motions in right ascension and declination) is ensured typically within 5.0--10 km\,s$^{-1}$ (see the bottom panel of Fig.\,1).

The standard Galactocentric cylindrical Coordinate ($R$, $\phi$, $Z$) is used in this paper, with the three direction velocities are $V_{R}$, $V_{\phi}$ and $V_{z}$, respectively.
To calculate the 3D positions and 3D velocities, we set the Galactocentric distance of the Sun as $R_{\odot}$\,=\,8.34\,kpc \citep{Reid2014}, Solar motions as ($U_{\odot}$, $V_{\odot}$, $W_{\odot}$)\,$=$\,$(13.00, 12.24, 7.24)$\,km\,s$^{-1}$ \citep{Schonrich2018} and, the local circular velocity as $V_{c,0}$\,=\,238.0\,km\,s$^{-1}$ \citep[e.g.,][]{Reid2004, Schonrich2010, Schonrich2012, Reid2014, Huang2015, Huang2016, Bland-Hawthorn2016}.
\begin{figure}[t]
\begin{center}
\includegraphics[scale=0.433,angle=0]{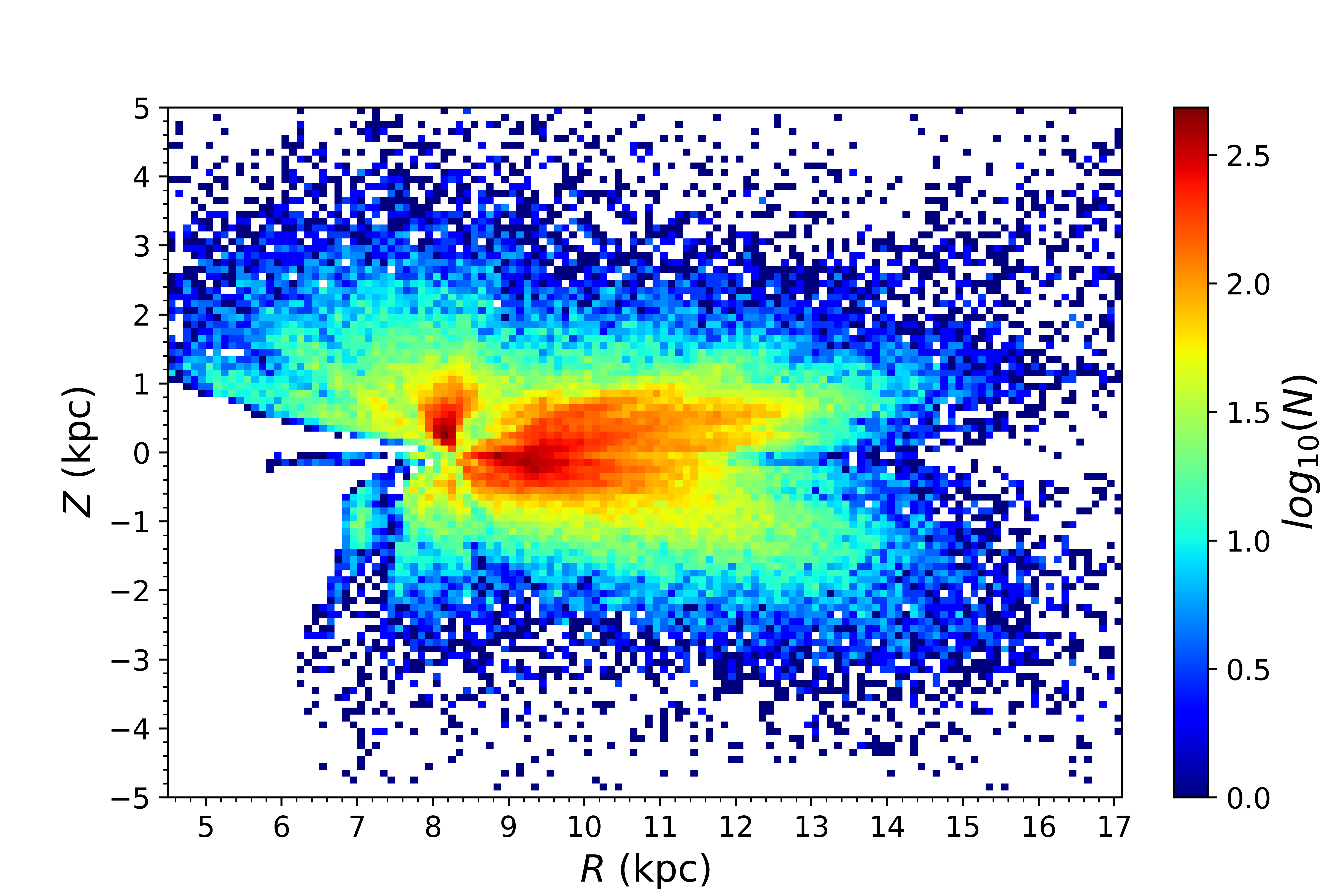}
\caption{The spatial distribution of the sample stars in the $R$ - $Z$ phase space, with both axes spaced by 0.1\,kpc, and color-code by stellar densities.}
\end{center}
\end{figure}

\begin{figure}[t]
\begin{center}
\includegraphics[scale=0.3,angle=0]{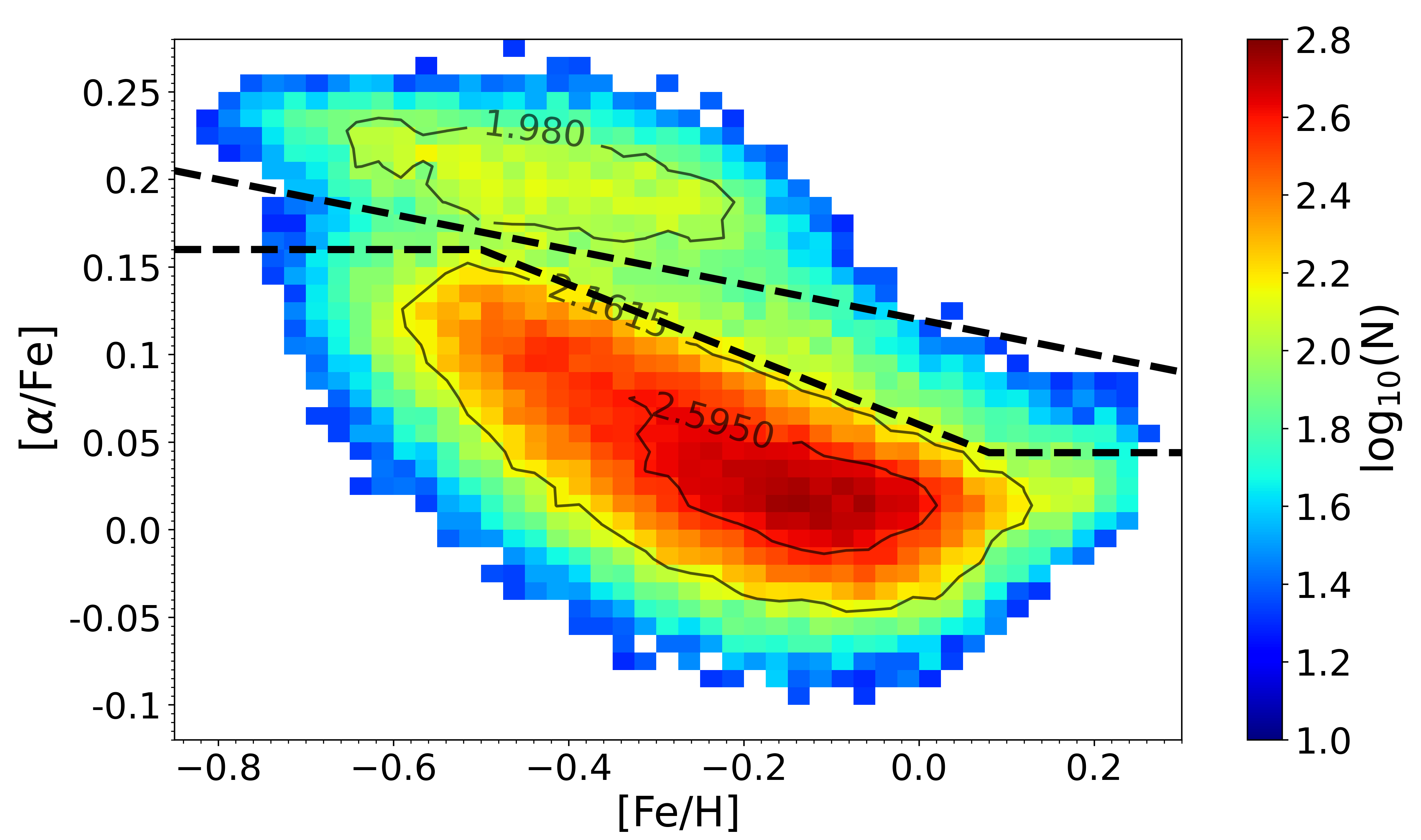}
\caption{The [Fe/H]$-$[$\alpha$/Fe] relation of the sample stars, color-coded by the stellar number densities.
The [Fe/H] and [$\alpha$/Fe] axes have respective widths of 0.025\,dex and 0.02\,dex.
There is a minimum of 20 stars in each bin.
The two lines are used to select the thick (above the lines) and the thin (below the lines) disks.}
\end{center}
\end{figure}

The velocity dispersions $\sigma$ of the individual spatial bin are calculated by 3$\sigma$-clipping produce that removes outliers.
To conduct a reliable Galactic disk stars analysis, we further set stellar vertical velocity\,$|V_{z}|$\,$\leq$\,120.0\,km\,s$^{-1}$ and [Fe/H] $\geq -1.0$ dex to exclude any possibility of halo stars.
Finally, 133,440 stars are eventually selected, and the spatial distribution of the final selected stars is shown in Fig.\,2.

\begin{table*}

\caption{The properties of thin/thin disk, mono-age and mono-[$\alpha$/Fe]-[Fe/H] populations.}

\centering
\setlength{\tabcolsep}{3mm}{
\resizebox{2.\columnwidth}{!}{
\begin{tabular}{lllllllll}
\hline
\hline
\specialrule{0em}{7pt}{0pt}
Name                                         &           [$\alpha$/Fe]            &        [Fe/H]      &     Mean R    &          $\sigma_{R}$            &       $\sigma_{Z}$  &  $\langle$ $V_{R}$ $V_{Z}$ $\rangle$ & Number     &    n\,(ratio) \\[0.07cm]
                                             &                (dex)               &        (dex)       &     (kpc)     &        (km\,s$^{-1}$)           &    (km\,s$^{-1}$)    &  (km$^{2}$\,s$^{-2}$)    &      &               \\
\specialrule{0em}{7pt}{0pt}
\hline
\specialrule{0em}{7pt}{0pt}
Thin disk                                    &                 -                  &         -          &     10.24     &            34.75               &          19.69       &  9.93 & 104,935 &       78.64\%  \\[0.3cm]
Thick disk                                   &                 -                  &         -          &     8.78      &            66.33               &          41.06       &  124.67 &  18,566 &       13.91\%  \\
\specialrule{0em}{7pt}{0pt}
\hline 
\specialrule{0em}{7pt}{0pt}
0.0 $<$ age $\leq$ 6.0\,Gyr (Y-Age)           &                -                   &         -          &     10.18     &            33.75               &          17.89       &  11.19 & 81,338  &      60.95\%  \\[0.3cm]
6.0 $<$ age $\leq$ 14.0\,Gyr (O-Age)          &                -                   &         -          &     9.71      &            51.13               &          32.15       &  22.18  & 52,102  &      39.05\%  \\

\specialrule{0em}{7pt}{0pt}
\hline
\specialrule{0em}{7pt}{0pt}

Low [$\alpha$/Fe] metal-poor (L$\alpha$MP)   & $-$0.3 $\leq$ [$\alpha$/Fe] $<$ 0.15 & [Fe/H] $<$ $-$0.2   &     10.72     &            34.68               &          21.68  &  15.83   &   60,295  &      45.19\%  \\[0.3cm]
Low [$\alpha$/Fe] metal-rich (L$\alpha$MR)   & $-$0.3 $\leq$ [$\alpha$/Fe] $<$ 0.15 & [Fe/H] $\geq$ $-$0.2 &     9.52      &            36.38               &          18.18  &  4.02   &   52,820  &      39.58\%  \\[0.3cm]
High [$\alpha$/Fe] metal-poor (H$\alpha$MP)  & 0.15 $\leq$ [$\alpha$/Fe] $<$ 0.5  & [Fe/H] $<$ $-$0.4   &     9.25      &            68.23               &          42.80   & 127.37   &   12,524  &      9.39\%   \\[0.3cm]
High [$\alpha$/Fe] metal-poor (H$\alpha$MR)  & 0.15 $\leq$ [$\alpha$/Fe] $<$ 0.5  & [Fe/H] $\geq$ $-$0.4 &     8.75      &            59.79               &          37.67   &  88.12  &   7,801   &      5.85\%   \\

\specialrule{0em}{7pt}{0pt}
\hline
\specialrule{0em}{7pt}{0pt}
\end{tabular}}
}
\label{tab:datasets}
\end{table*}

\begin{figure*}[t]
\centering

\subfigure{
\includegraphics[width=17cm]{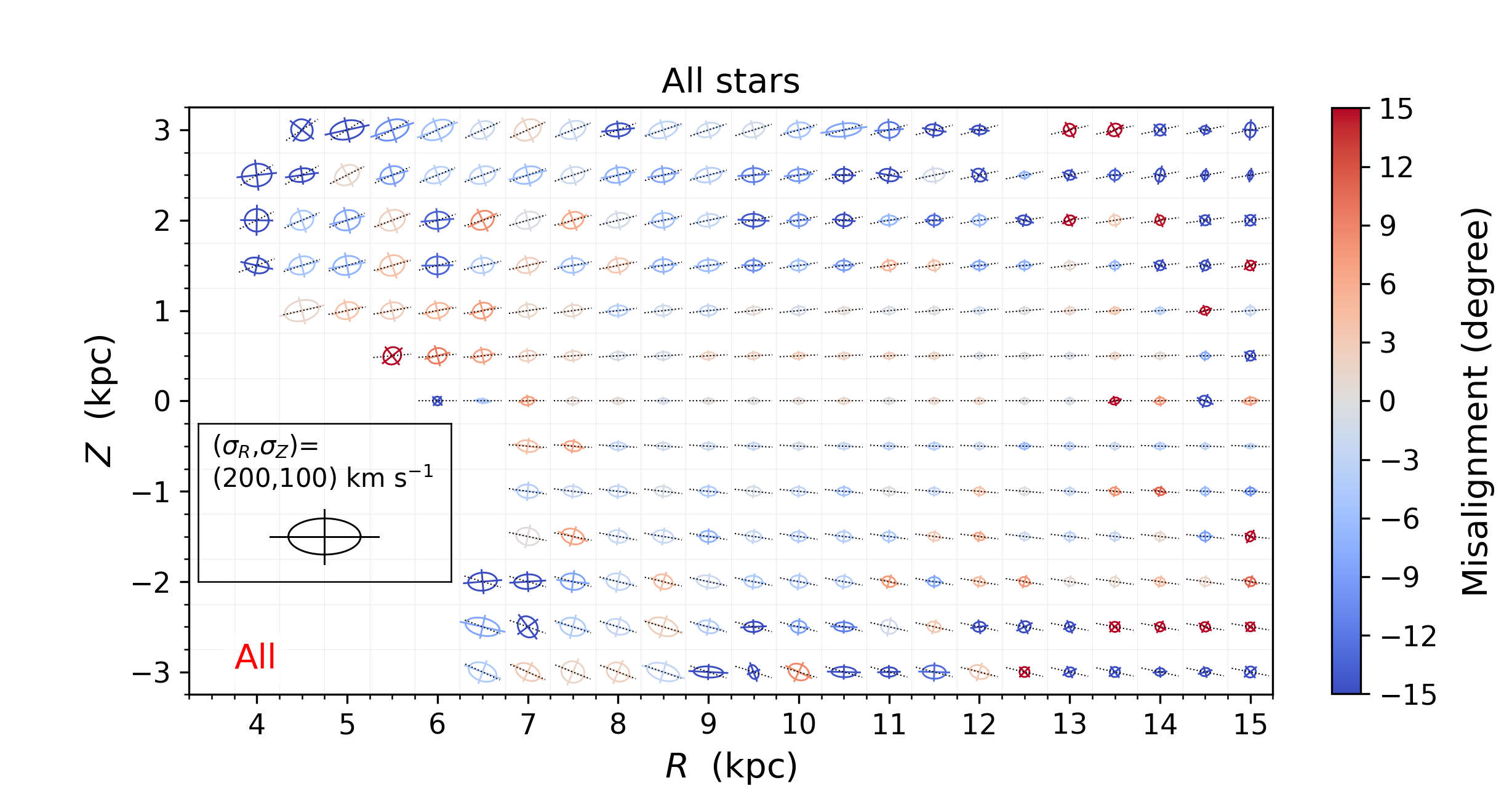}
}

\caption{Velocity ellipsoid distribution in $R$--$Z$ phase space for all the RC sample stars, with bins of size 0.5\,kpc for both axes, and no less than 10 stars in each bin.
The velocity ellipses are color-coded by their misalignment with respect to spherical alignment.
The orientation that corresponds to spherical alignment is indicated by the dotted grey line through each ellipse.
The size of the ellipsoid is proportional to the value of the velocity dispersion in each bin.
The black rectangle inset in the figure shows the velocity ellipse for a non-tilted distribution with dispersions with ($\sigma_{R}$,\,$\sigma_{Z}$)=(200,\,100)\,km\,s$^{-1}$.}
\end{figure*}

\begin{figure*}[t]
\centering

\subfigure{
\includegraphics[width=17cm]{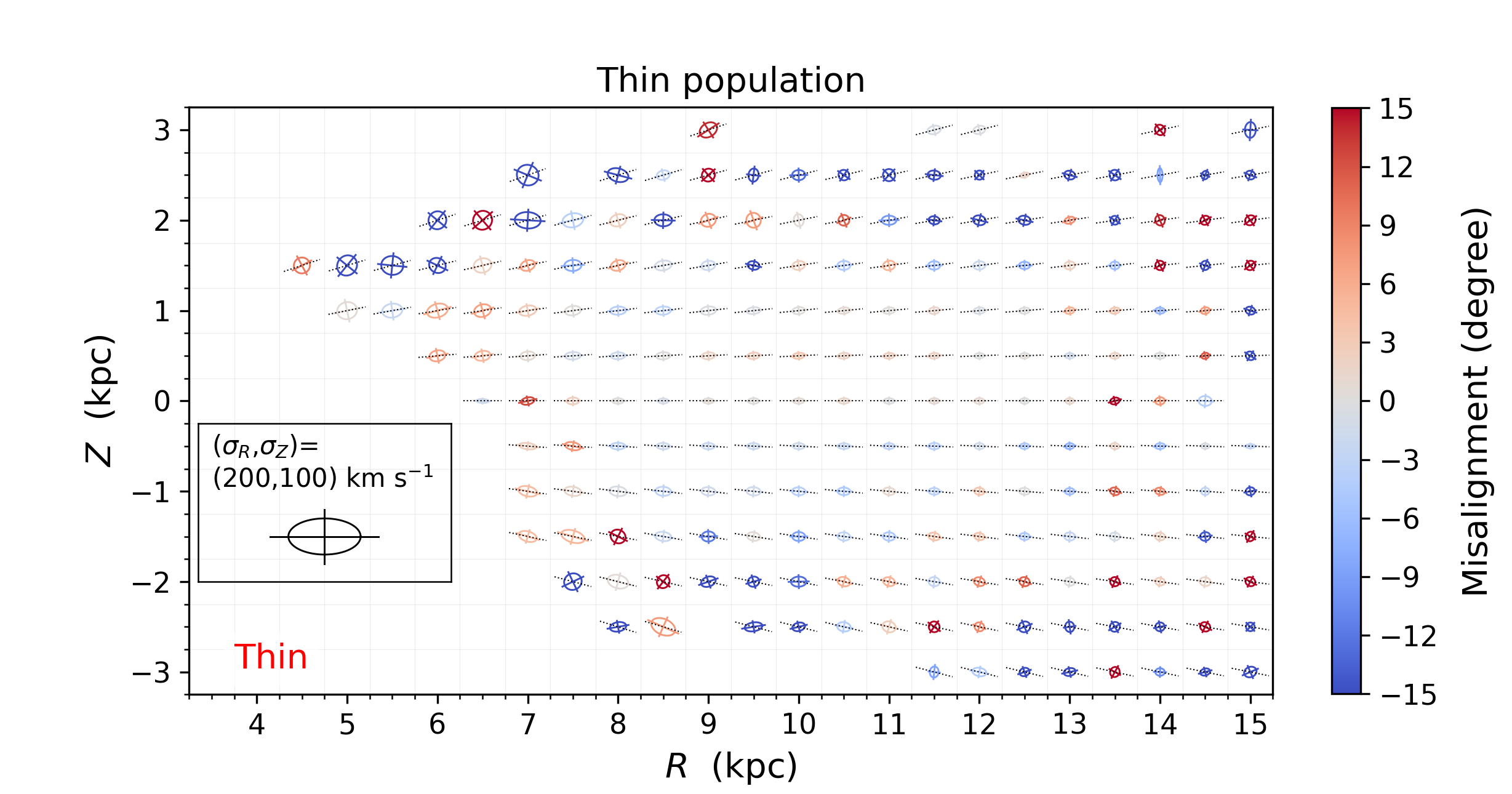}
}

\subfigure{
\includegraphics[width=17cm]{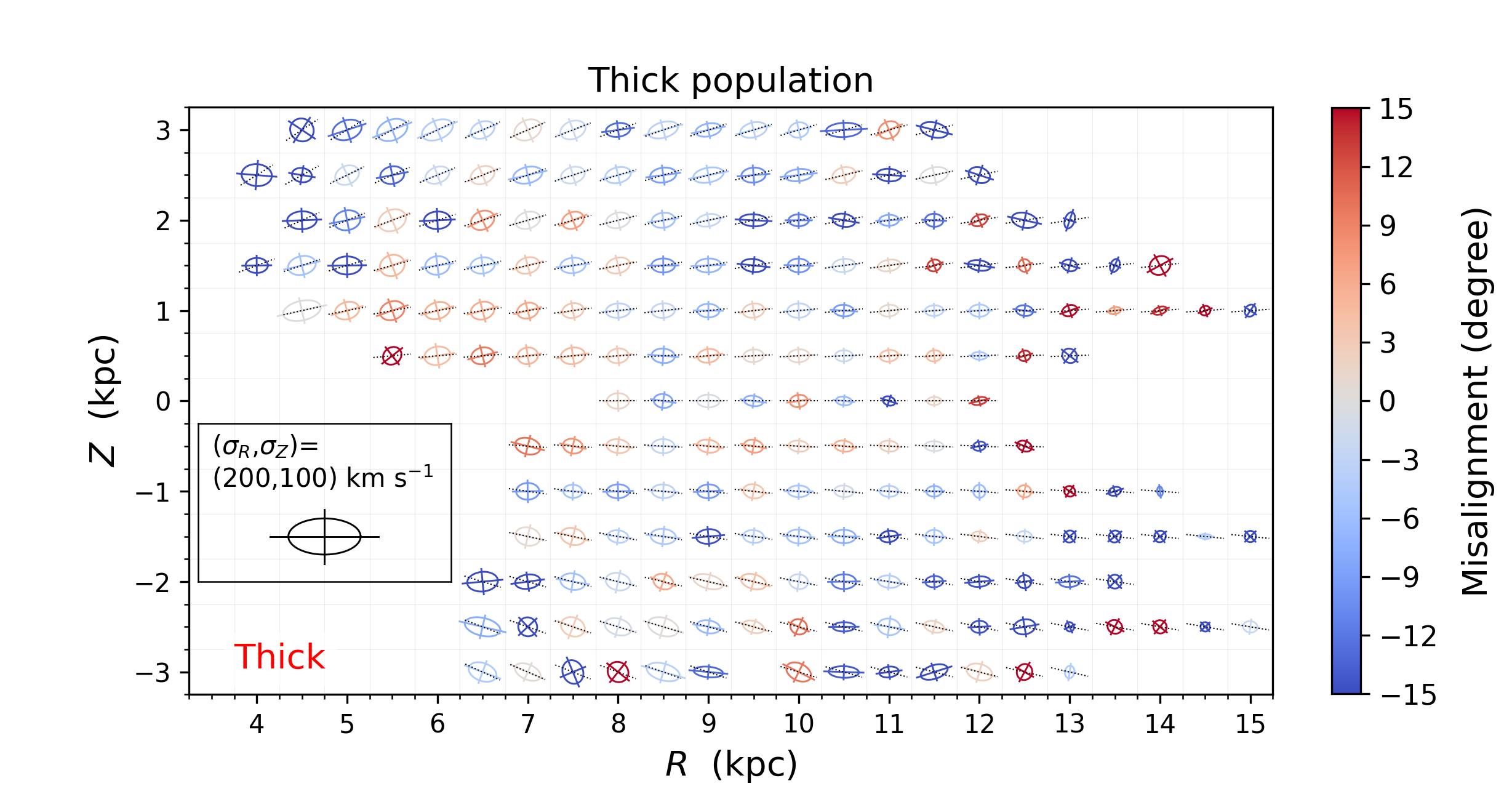}
}

\caption{Similar to Fig.\,4 but for the thin and thick disks.}
\end{figure*}

\section{Method}

We calculate the orientation of the velocity ellipse as follows Smith et al. ({\color{blue}{2009}}), that is, for two orthogonal velocity components, $v_{i}$, $v_{j}$, the tilt angle $\alpha_{ij}$ defined as:

\begin{equation}
\label{eq:tiltangle1}
    \tan(2 \alpha_{ij}) = \frac{2 \mathrm{cov}(v_i, v_j)}{\mathrm{var}(v_i) - \mathrm{var}(v_j)} = \frac{2\sigma_{ij}^{2}}{\sigma_{ii}^{2} - \sigma_{jj}^{2}}
\end{equation}
where

\begin{equation}
\mathrm{cov}(v_i, v_j) \equiv \sigma^2_{ij} \equiv
\left\langle
(v_i - \langle v_i \rangle)(v_j - \langle v_j \rangle)
\right\rangle
\label{eq:cov}
\end{equation}
and the angled brackets are the averaging of the phase-space distribution function \citep[see e.g.,][]{Binney2008, Smith2009, Evans2016}.

The tilt angle, $\alpha_{ij}$, is the angle that is measured from the i-axis to the major axis of the ellipse formed by projecting the three-dimensional velocity ellipsoid onto the ij-plane \citep[e.g.,][]{Binney1998}, with values of counter-clockwise measurement is positive, and hence takes values from $-$45$^{\circ}$ to $+$45$^{\circ}$.
In the Galactocentric cylindrical coordinates ($R$, $\phi$, $Z$), when the tilt angle $\alpha$ = 0$^{\circ}$, which implies the major (direction of $V_{R}$) and minor (direction of $V_{Z}$) axis of the velocity ellipsoid are respectively exactly aligned with R-axis and Z-axis.

To make a more detailed analysis of the tilt of the velocity ellipsoid,  we further separate our sample into the thin/thick disk, mono-age and mono-[$\alpha$/Fe]-[Fe/H] populations.
The separation of the thin and thick disks is shown in Fig.\,3.
We used two cuts to separate the two populations with combined previous studies \citep[e.g.,][]{Bensby2014, Lee2011, Brook2012, Haywood2013, Recio-Blanco2014, Guiglion2015}, and then, selected low-[$\alpha$/Fe] with 104,935 stars as thin disk population and, high-[$\alpha$/Fe] with 18,566 stars as thick disk population, respectively.

For the definition of mono-age, we separate our sample into young age population (with age\,$\leq$\,6.0\,kpc, hereafter the Y-Age population) and old age population (with age\,$>$\,6.0\,Gyr, hereafter the O-Age population), with separate the so-called ``young" [$\alpha$/Fe]-enhanced stars into old populations since they are genuinely old \citep[e.g.,][]{Chiappini2015, Jofre2016, Sun2020}.
The properties of various mono-age populations are shown in Table\,1.
For the definition of mono-[$\alpha$/Fe]-[Fe/H] populations, we divide the RC sample stars into high-[$\alpha$/Fe] population with 0.15\,$\leq$\,[$\alpha$/Fe]\,$<$\,0.5\,dex, and low-[$\alpha$/Fe] population with $-$0.3\,$\leq$\,[$\alpha$/Fe]\,$<$\,0.15\,dex, and we further separate each [$\alpha$/Fe] population into metal-rich and metal-poor sub-populations by considering the almost comparable number of stars in each sub-population, and the results are also shown in Table\,1.

\begin{table*}
\caption{The $\alpha_{0}$ of various populations.}

\centering
\setlength{\tabcolsep}{11mm}{
\begin{tabular}{lllllllll}
\hline
\hline
\specialrule{0em}{7pt}{0pt}
Name                                         &     &          $\alpha_{0}$       \\
\specialrule{0em}{7pt}{0pt}
\hline
\specialrule{0em}{7pt}{0pt}
All stars                                    &     &        0.68 $\pm$ 0.05 \\[0.3cm]
Thin disk                                    &     &        0.72 $\pm$ 0.08 \\[0.3cm]
Thick disk                                   &     &        0.64 $\pm$ 0.06 \\
\specialrule{0em}{7pt}{0pt}
\hline 
\specialrule{0em}{7pt}{0pt}
Y-Age           &      &   0.69 $\pm$ 0.07   \\[0.3cm]
O-Age           &      &   0.65 $\pm$ 0.06   \\
\specialrule{0em}{7pt}{0pt}
\hline
\specialrule{0em}{7pt}{0pt}
L$\alpha$MP    &     &  0.72 $\pm$ 0.10  \\[0.3cm]
L$\alpha$MR    &     &  0.85 $\pm$ 0.08   \\[0.3cm]
H$\alpha$MP    &     &  0.61 $\pm$ 0.07   \\[0.3cm]
H$\alpha$MR    &     &  0.75 $\pm$ 0.10   \\
\specialrule{0em}{7pt}{0pt}
\hline
\specialrule{0em}{7pt}{0pt}
\end{tabular}}
\label{tab:datasets}
\end{table*}

\section{The velocity ellipsoid of the thin/thick disk, mono-age and mono-[$\alpha$/Fe]-[Fe/H] populations}

\begin{figure*}[t]
\centering

\subfigure{
\includegraphics[width=15.5cm]{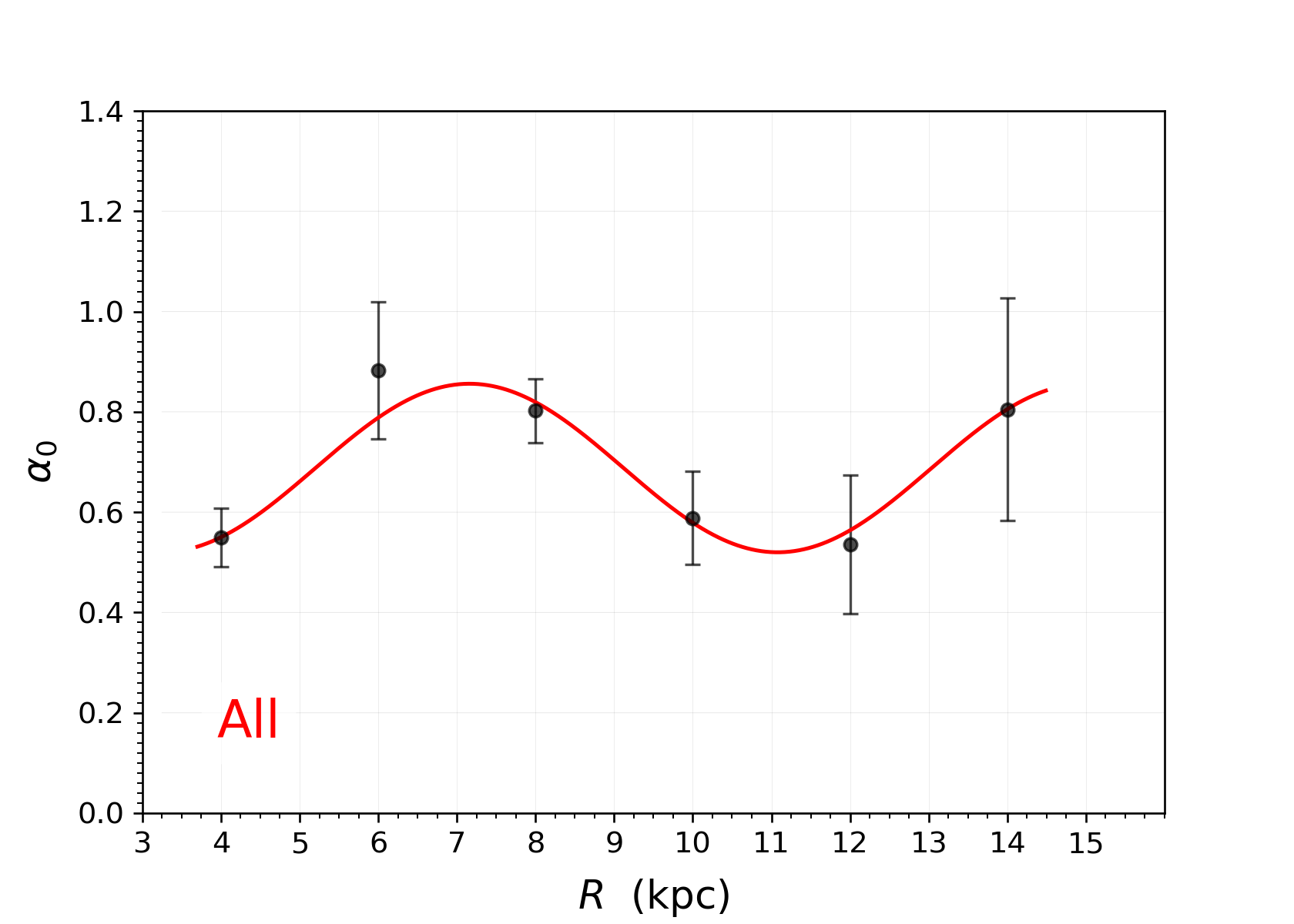}
}

\caption{The $\alpha_{0}$ as a function of $R$ for all the RC sample stars.
The red line is the best fit with equation (4).}
\end{figure*}

The results of the velocity ellipsoid for various populations are shown in Fig.\,4--5 and Appendix Fig.\,A1, with color-coded by the misalignment with respect to spherical alignment.
The misalignment is defined as $\alpha$ $-$ $\alpha_{sph}$ for $Z$ $\geq$ 0.0\,kpc, and as $\alpha_{sph}$ $-$ $\alpha$ for $Z$ $<$ 0.0\,kpc, where $\alpha$ is the tilt angle of velocity ellipsoid and $\alpha_{sph}$ is the angle from R-major to the orientation of a spherically aligned velocity dispersion.
The red and blue colors are respectively corresponding to steeper tilt angles and shallower tilt angles.

For all the RC sample stars (Fig.\,4), the spherical alignment at $R$ $\sim$ 6.0\,kpc tends to cylindrical alignment with $R$ increases at $R$ $\leq$ 12.0\,kpc, and then again tends to spherical alignment outwards.
The trend of the velocity ellipsoids distributions with Galactic height is distinguished for the inner and outer disk regions.
For $R$ $\leq$ 11.5\,kpc, the velocity ellipsoid of higher $|Z|$ tends to relatively cylindrical alignment, while for $R$ $>$ 11.5\,kpc, it is relatively spherical arrangement with increasing $|Z|$.
Stars with larger velocity dispersion tend to have a globally tilted towards cylindrical arrangement of the velocity ellipsoid.

For the thin disk stars (see Fig.\,5), the velocity dispersions show a weak increasing trend at $R$ $\geq$ 12.5\,kpc, which indicates the thin disk warp.
In the inner disk region, stars with the largest velocity dispersion generally have a tilted towards cylindrical arrangement of the velocity ellipsoid.
While for the outer disk region, stars with the largest velocity dispersion generally have a random alignment of the velocity ellipsoid.

For the thick disk stars (see Fig.\,5), the velocity dispersions display no obvious global trend with $R$ and $Z$, but stars with the largest Galactic height generally have a tilted towards cylindrical arrangement of the velocity ellipsoid.
The velocity ellipsoids of the thick disk stars tend to be more randomly aligned than the thin disk stars, with the spherical alignment at lower Galactic height tending to cylindrical alignment towards higher Galactic height, while for the radial direction, it is not a global trend.

For our different age and [$\alpha$/Fe]-[Fe/H] populations (see Appendix Fig.\,A1), the results display that the Y-Age, L$\alpha$MP and L$\alpha$MR populations show the similar result to the thin disk stars, while the O-Age, H$\alpha$MP and H$\alpha$MR display a thick disk like result.
The velocity dispersions of the young/low-[$\alpha$/Fe] populations (left panels) are globally smaller than old/high-[$\alpha$/Fe] populations (right panels).
The velocity ellipsoids of old/high-[$\alpha$/Fe] populations tend to be more randomly aligned than young/low-[$\alpha$/Fe] populations.
This randomly aligned may be strongly related to the different dynamical heating histories of those populations.

\begin{table*}
\caption{Parameters of $\alpha_{0}$ = $A$\,$\mathrm{sin}$($\omega R$ + $\Phi$) + $C$ obtained for different populations.}

\centering
\setlength{\tabcolsep}{11mm}{
\begin{tabular}{lllllllll}
\hline
\hline
\specialrule{0em}{7pt}{0pt}
Name                                &  $A$   &  $\omega$ & $\Phi$  &  $C$       \\
\specialrule{0em}{7pt}{0pt}
\hline
\specialrule{0em}{7pt}{0pt} 
All stars                           &   0.17\,$\pm$\,0.06    &    0.80\,$\pm$\,0.09       &  2.11\,$\pm$\,0.63       &   0.69\,$\pm$\,0.04        \\[0.3cm]
Thin disk                           &   0.23\,$\pm$\,0.11    &    0.80\,$\pm$\,0.09       &  2.47\,$\pm$\,0.53       &   0.81\,$\pm$\,0.08        \\[0.3cm]
Thick disk                          &   0.21\,$\pm$\,0.09    &    0.82\,$\pm$\,0.09       &  2.28\,$\pm$\,0.58       &   0.69\,$\pm$\,0.06         \\
\specialrule{0em}{7pt}{0pt}
\hline 
\specialrule{0em}{7pt}{0pt}
Y-Age                               &   0.19\,$\pm$\,0.06    &    0.87\,$\pm$\,0.08       &  2.17\,$\pm$\,0.64       &   0.66\,$\pm$\,0.07        \\[0.3cm]
O-Age                               &   0.18\,$\pm$\,0.07    &    0.84\,$\pm$\,0.08       &  2.15\,$\pm$\,0.62       &   0.73\,$\pm$\,0.05         \\
\specialrule{0em}{7pt}{0pt}
\hline
\specialrule{0em}{7pt}{0pt}
L$\alpha$MP                         &   0.23\,$\pm$\,0.12    &    0.78\,$\pm$\,0.08       &  2.11\,$\pm$\,0.66       &   0.77\,$\pm$\,0.09        \\[0.3cm]
L$\alpha$MR                         &   0.41\,$\pm$\,0.09    &    0.88\,$\pm$\,0.04       &  2.71\,$\pm$\,0.32       &   0.88\,$\pm$\,0.07        \\[0.3cm]
H$\alpha$MP                         &   0.06\,$\pm$\,0.06    &    0.75\,$\pm$\,0.18       &  2.12\,$\pm$\,0.66       &   0.63\,$\pm$\,0.06        \\[0.3cm]
H$\alpha$MR                         &   0.21\,$\pm$\,0.12    &    0.86\,$\pm$\,0.09       &  2.36\,$\pm$\,0.56       &   0.88\,$\pm$\,0.08         \\
\specialrule{0em}{7pt}{0pt}
\hline
\specialrule{0em}{7pt}{0pt}
\end{tabular}}
\label{tab:datasets}
\end{table*}

\section{The tilt of the velocity ellipsoid for various populations}

Previous studies provide a simple model to summarize the results of the tilt angle of the velocity ellipsoids \citep[e.g.,][]{Binney2014, Budenbender2015}, which suggested $\alpha$ followed as

\begin{equation}
\label{eq:tiltangle2}
    \alpha = \alpha_{0}\, \mathrm{arctan}(\frac{Z}{R})
\end{equation}
where $\alpha_{0}$ is the fitting constant, and a result of $\alpha_{0}$\,=\,1.0 means exact spherical alignment, whilst $\alpha_{0}$\,$<$\,1.0 implies the ellipsoids tend to tilt towards cylindrical alignment.

We fit the results using the emcee python package (Foreman-Mackey et al. {\color{blue}{2013}}) on all bins with no less than 10 stars since these bins still contain much valuable information about the ellipsoid alignment, although these bins still have larger uncertainties, and bins with fewer stars tend to randomly aligned.
The fitting result for various populations is summarized in Table\, 2.

Our results indicate that the tilt of all the RC sample stars values $\alpha_{0}$ = 0.68\,$\pm$\,0.05, which is consistent with a tilted towards cylindrical alignment.
This result is slightly smaller than the local RAVE data of Binney et al. ({\color{blue}{2014}}), who reported $\alpha_{0}$\,$\sim$\,0.8.
It is in obvious disagreement with $\alpha_{0}$\,=\,0.90\,$\pm$\,0.04 determined in B{\"u}denbender et al. ({\color{blue}{2015}}) and $\alpha_{0}$\,=\,0.95\,$\pm$\,0.01 determined in Everall et al. ({\color{blue}{2019}}).
It is worth emphasizing that we also remove stars by using 3$\sigma$-clipping of each velocity distribution for each bin, and any possible halo stars are also excluded from our results.
We further fit the model on bins with no less than 20 stars, the result indicates that $\alpha_{0}$ = 0.72\,$\pm$\,0.05, which is still significantly smaller than previous results on samples containing halo stars \citep[e.g.,][]{Binney2014, Budenbender2015, Everall2019, Hagen2019}, implying that any deviation from the spherical alignment of the velocity ellipsoids may be caused by the gravitational potential of the baryonic disk.

For our different populations, the results show that they are almost tilted towards cylindrical in alignment, meaning that all populations are likely significantly affected by the gravitational potential of the baryonic disk of the Galaxy.
This is an acceptable result since the spatial distributions of all populations are concentrated on the Galactic disk, which is also the reason why our $\alpha_{0}$ is generally smaller than other results \citep[e.g.,][]{Budenbender2015, Everall2019, Hagen2019}.
The study of the alignment of the velocity dispersion tensor revealing the underlying gravitational potential \citep[e.g.,][]{An2016, Evans2016}.
The results indicate that if the alignment is everywhere spherical, then the potential must be spherical (if non-singular).
If the alignment is everywhere cylindrical, the potential must be separable in cylindrical coordinates.
In our results, the alignment is neither spherical nor cylindrical, but in-between.
Such coupling between radial and vertical motion captured in the velocity ellipsoid tilt tells us about the interplay between the disc and halo potentials.

The result of the L$\alpha$MR indicates that $\alpha_{0}$\,=\,0.85\,$\pm$\,0.08.
Although this result is in good agreement with previous results \citep[e.g.,][]{Binney2014, Budenbender2015, Everall2019}, it is obviously larger than other populations, meaning that the L$\alpha$MR tends to be more spherically aligned than other populations.
Some possible reasons may explain this behavior;
(i) larger uncertainties of this population.
However, the results with no less than 20 stars per bin of L$\alpha$MR indicate that $\alpha_{0}$\,=\,0.82\,$\pm$\,0.08.
This may mean that the effect of uncertainties on the results is not obvious;
(ii) natural property of this population, stars of L$\alpha$MR are generally young, and their kinematic properties are similar to the ISM that they are born.

We do observe a slight age and [$\alpha$/Fe] dependence of the tilt of the velocity ellipsoids, with the $\alpha_{0}$ for young/low-[$\alpha$/Fe] populations are larger than old/high-[$\alpha$/Fe] population.
In detail, the $\alpha_{0}$ = 0.72\,$\pm$\,0.08 for the thin disk stars is larger than $\alpha_{0}$\,=\,0.64\,$\pm$\,0.06 for the thick disk stars.
And the $\alpha_{0}$ for Y-Age, L$\alpha$MP and H$\alpha$MP populations are respectively larger than that for O-Age, L$\alpha$MR and H$\alpha$MR populations.
These may imply that young/low-[$\alpha$/Fe] stars tend to be spherically aligned than old/high-[$\alpha$/Fe] stars, and this translates to the effect of the gravitational potential of the baryonic disk increases with stellar age (or [$\alpha$/Fe]) increases.
This is likely because stars with old age are generally kinematically relaxed, and therefore, their velocity ellipsoids are mainly dominated by the gravitational potential of the baryonic disk.

Using a sample from Gaia Radial Velocity Spectrometer, Everall et al. ({\color{blue}{2019}}) reported that the tilt of the velocity ellipsoid displays a strong correlation with spatial positions.
In their results, for stars with $|Z|$ $<$ 2.0\,kpc, $\alpha_{0}$ = 0.950 $\pm$ 0.007, whilst for stars with $|Z|$ $>$ 2.0\,kpc, $\alpha_{0}$ = 0.966 $\pm$ 0.018.
For stars with $R$ $<$ 7.0\,kpc, $\alpha_{0}$ = 0.917 $\pm$ 0.013, whilst for stars with $R$ $>$ 7.0\,kpc, $\alpha_{0}$ = 0.963 $\pm$ 0.007.
Similar results can be also found in Hagen et al. ({\color{blue}{2019}}), who suggested that the tilt angle of the velocity ellipsoid in spherical coordinates as $\alpha$(r, $Z$) $\sim$ 0.72\,($r$\,$-$\,6.16)\,$Z$.

In Fig.\,6, we present the $\alpha_{0}$ as a function of $R$ for all the RC sample stars.
The result indicates that the $\alpha_{0}$ shows oscillations with $R$.
At $R$ = 8.0\,kpc, the $\alpha_{0}$ $\sim$ 0.81 $\pm$ 0.06, which is in good agreement with the result of $\alpha_{0}$ $\sim$ 0.80 in Binney et al. ({\color{blue}{2014}}) with the local RAVE data.
At $R$ = 6.0\,kpc, the $\alpha_{0}$ $\sim$ 0.89 $\pm$ 0.13 is also well consistent with $\alpha_{0}$ $\sim$ 0.90 $\pm$ 0.04 in the solar neighbourhood determined by B{\"u}denbender et al. ({\color{blue}{2015}}).
The profile of the $\alpha_{0}$ -- $R$ displays a significant sine-function shape, and therefore, we fit the profile with a sine-function as follows:

\begin{equation}
\label{eq:tiltangle2}
    \alpha_{0} = A\,\mathrm{sin}(\omega R + \Phi) + C
\end{equation}
The best fit as shown by the red line in the figure, the result yields $\alpha_{0}$ = 0.17\,$\mathrm{sin}$\,(0.80\,$R$\,+\,2.11)\,+\,0.69.
Based on the best fit, we can determine the $\alpha_{0}$ has a local maximum at R $\sim$ 7.18\,kpc with $\alpha_{0}$ $\sim$ 0.85, and a local minimum at R $\sim$ 11.12\,kpc with $\alpha_{0}$ $\sim$ 0.52.
Since our sample stars are mostly disk stars, the oscillations of the $\alpha_{0}$ with $R$ may be strongly linked to the structures of the disk, and therefore, the minimum of the $\alpha_{0}$ at R $\sim$ 11.12\,kpc is likely because of the Perseus arm.
And the Perseus arm traced by our analysis located at $R$ $\sim$ 11.12\,kpc is also in rough agreement with previous studies \citep[e.g.,][]{Chen2019, Sun2023}.

For our different populations, the oscillations can also be found for various populations (Appendix Fig.\,B1), with the thin disk, Y-Age and low-[$\alpha$/Fe] populations showing similar shapes, while the thick disk, O-Age and High-[$\alpha$/Fe] populations display another similar shape.
We also fit the $\alpha_{0}$ -- $R$ with equation (4) for various populations, with the best fit as shown by the red line in the figure, and the results are summarized in Table\,3.
The results indicate that the oscillations of $\alpha_{0}$ for young/low-[$\alpha$/Fe] populations (left panels in Appendix Fig.\,B1) are stronger than old/high-[$\alpha$/Fe] populations (right panels in Appendix Fig.\,B1).
Since the oscillations in $\alpha_{0}$ appear in both kinematically hot and cold populations, we can further confirm that one of its mechanisms is likely dynamical in origin.
The most likely culprit is resonances with the Galactic bar.
This would cause perturbations in the stellar orbits at locations of the principal resonances \citep[e.g.,][]{Collett1997} and has been seen in Hipparcos and Gaia data \citep[e.g.,][]{Dehnen2000, Trick2021}.

The north-south asymmetry is also obvious for the velocity ellipsoids, similar result is also reported by Everall et al. ({\color{blue}{2019}}).
Although the explanation of this phenomenon has not yet reached a unified conclusion, some properties may be linked to this phenomenon, for example, the substructure and streams of the Galactic disk, buckling of Galactic central bar \citep[e.g.,][]{Saha2013}, the effect of the disk bending \citep[e.g.,][]{Gomez2013, Williams2013, Xu2015, Laporte2019} flare \citep[e.g.,][]{Minchev2015, Bovy2016} and warp \citep[e.g.,][]{Momany2006, Williams2013}, or some unrecognized systematics in the data.

\section{Conclusions}

In this paper, we use nearly 140,000 RC stars selected from the LAMOST and {\it Gaia} to calculate the tilt of the velocity ellipsoid for thin/thick disks, mono-age and mono-[$\alpha$/Fe]-[Fe/H] populations over a wider range of Galactocentric radii with higher accuracy.
We find that the tilt angles of the velocity ellipsoid are distinguished for various populations, and the tilt angles display a strong correlation with age and population.

The sample with 137,448 RC stars is tilted towards cylindrical alignment.
The tilt of all the RC sample stars is accurately described as $\alpha$ = (0.68 $\pm$ 0.05)\,$\mathrm{arctan}$\,($Z$/$R$), which is generally smaller than the previous results on samples including halo stars \citep[e.g.,][]{Budenbender2015, Everall2019, Hagen2019}.
Since our RC sample stars are dominated by the disk stars, our result reveals that any deviation from the spherical alignment of the velocity ellipsoids is likely because of the gravitational potential of the baryonic disk of the Galaxy.

The thin and thick disks values $\alpha$ = (0.72 $\pm$ 0.08)\,$\mathrm{arctan}$\,($Z$/$R$) and $\alpha$ = (0.64 $\pm$ 0.07)\,$\mathrm{arctan}$\,($Z$/$R$), respectively.
The $\alpha_{0}$ displays an age (and [$\alpha$/Fe]) dependence, with a decreasing trend in $\alpha_{0}$ as age (and [$\alpha$/Fe]) increases, meaning that the velocity ellipsoids of the kinematically relaxed stars are mainly dominated by the gravitational potential of the baryonic disk.

Our detailed analysis of the correlation of $\alpha_{0}$ and $R$ of various populations indicates that the $\alpha_{0}$ displays oscillations with $R$ for all the different populations.
For all the RC sample stars is $\alpha_{0}$ = 0.17\,$\mathrm{sin}$\,(0.80\,$R$\,+\,2.11)\,+\,0.69.
For the thin and thick disks are $\alpha_{0}$ = 0.23\,$\mathrm{sin}$\,(0.80\,$R$\,+\,2.47)\,+\,0.81 and $\alpha_{0}$ = 0.21\,$\mathrm{sin}$\,(0.82\,$R$\,+\,2.28)\,+\,0.69, respectively.
The oscillations of $\alpha_{0}$ for both kinematically hot and cold populations, which is likely because of the resonances with the Galactic bar.

\section*{Acknowledgements}

Guoshoujing Telescope (the Large Sky Area Multi-Object Fiber Spectroscopic Telescope LAMOST) is a National Major Scientific Project built by the Chinese Academy of Sciences. Funding for the project has been provided by the National Development and Reform Commission. LAMOST is operated and managed by the National Astronomical Observatories, Chinese Academy of Sciences. The LAMOST FELLOWSHIP is supported by Special Funding for Advanced Users, budgeted and administrated by Center for Astronomical Mega-Science, Chinese Academy of Sciences (CAMS)

It is a pleasure to thank Yang Huang for providing assistance in revising the manuscript.

This work is supported by National Key R\&D Program of China No. 2019YFA0405500 and National Natural Science Foundation of China grants 11903027, 11973001, 11833006, 11811530289, U1731108, U1531244, 11573061, 11733008, 12073070, and 12003027.

\appendix

\section{Velocity ellipsoid distribution for various mono-age and mono-[$\alpha$/Fe]-[Fe/H] populations}

Fig. {\color{blue}{A1}} presents the distributions of the velocity ellipsoid for various mono-age and mono-[$\alpha$/Fe]-[Fe/H] populations.

\begin{figure*}[t]
\centering
\subfigure{
\includegraphics[width=17cm]{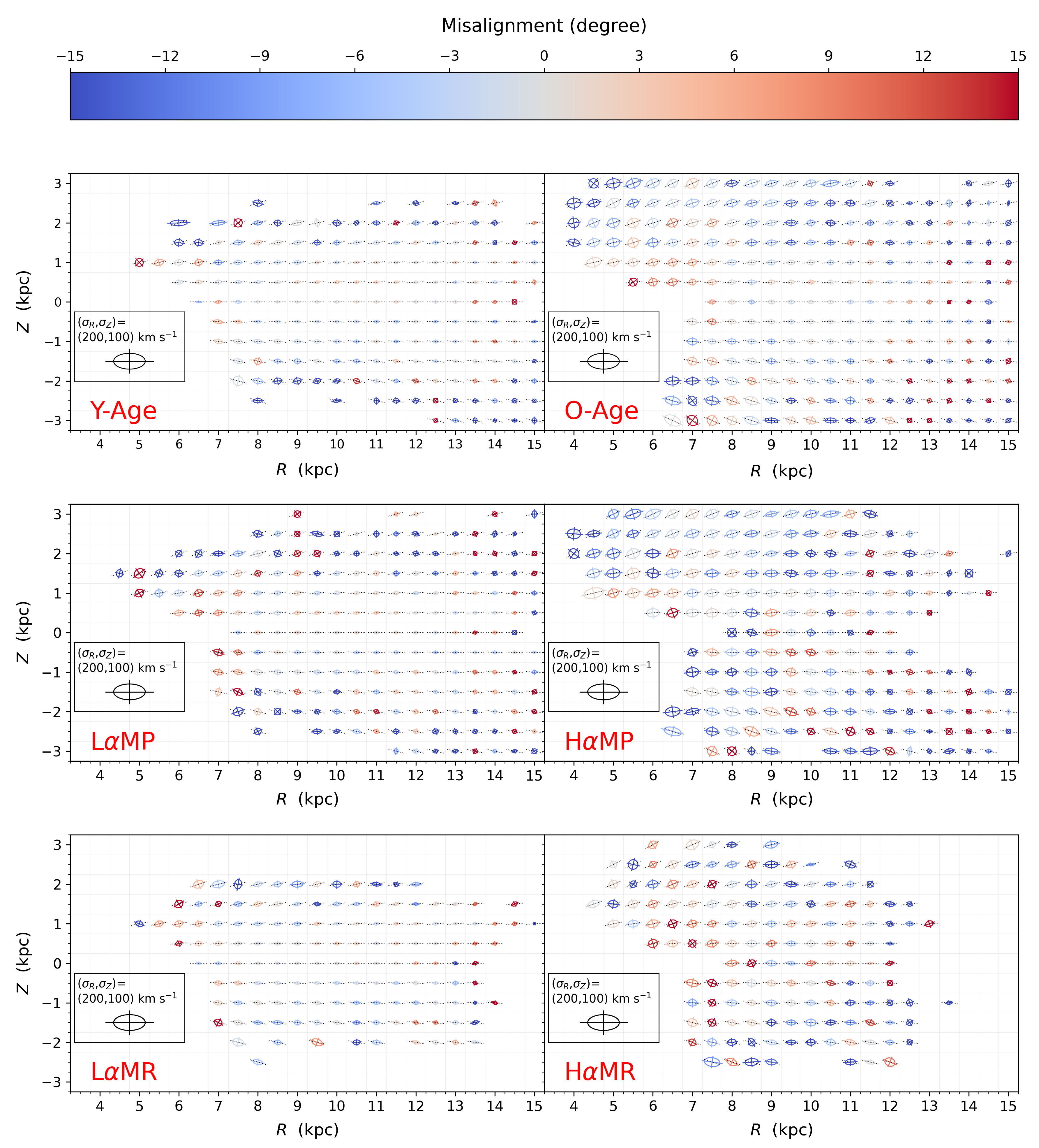}
}
\caption{Velocity ellipsoid distribution for various mono-age and mono-[$\alpha$/Fe]-[Fe/H] populations, with both axes are spaced by 0.5\,kpc.
There is minimum 10 stars in per bin.}
\end{figure*}

\section{The $\alpha_{0}$ as function of $R$ for various populations}

Fig. {\color{blue}{B1}} presents the $\alpha_{0}$ as function of $R$ for the thin/thick disk, mono-age and mono-[$\alpha$/Fe]-[Fe/H] populations.
The red lines are the best fit with equation (4) for the data.

\begin{figure*}[t]
\centering

\subfigure{
\includegraphics[width=17cm]{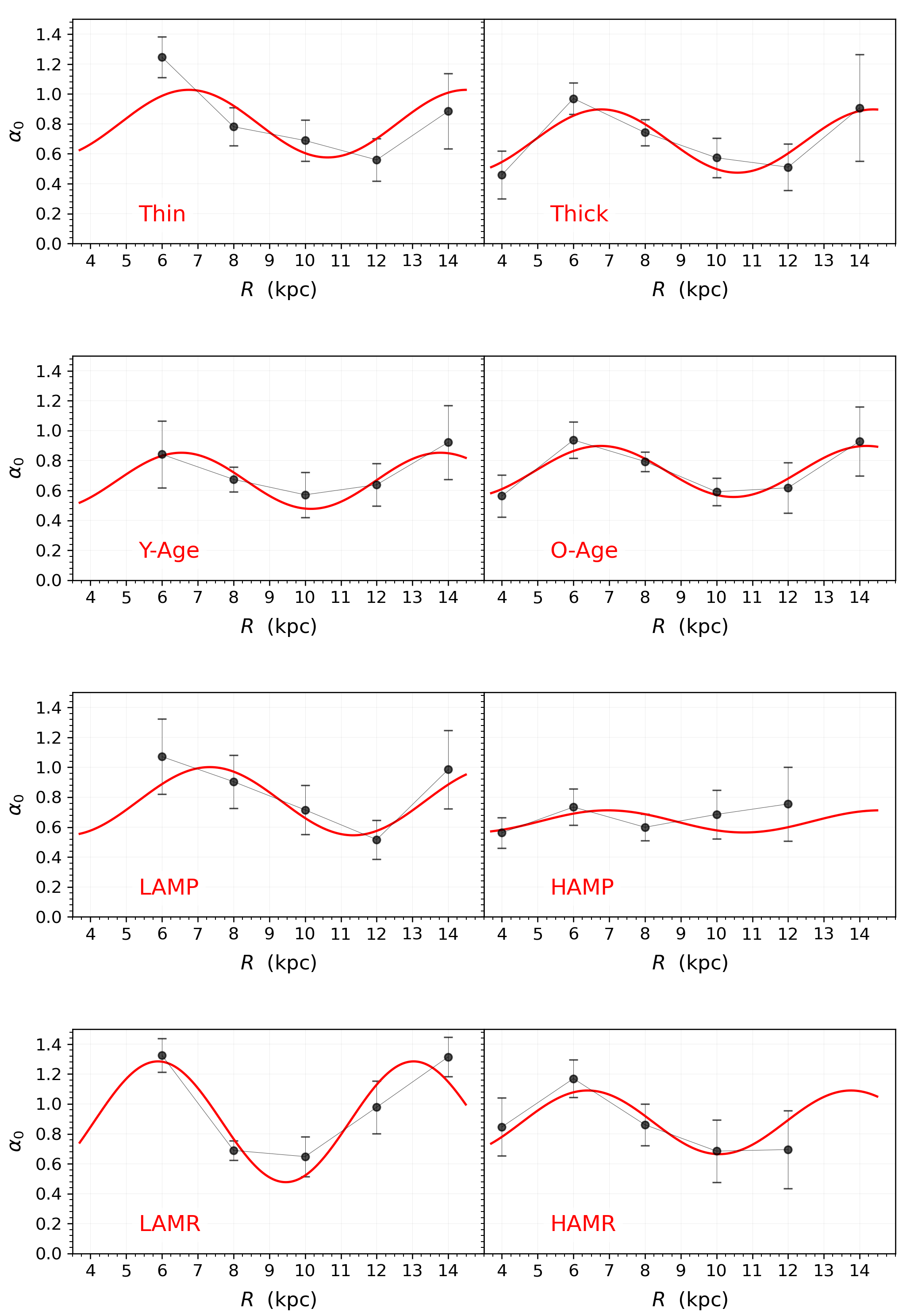}
}

\caption{The $\alpha_{0}$ as function of $R$ for various populations.
The red lines are the best fit with equation (4) for the data.}
\end{figure*}


\begin{thebibliography}{}


\bibitem[Ablimit et al.(2020)]{Ablimit2020} Ablimit, I., Zhao, G., Flynn, C., Bird, S. A.\ 2020, \apj, 895, 12

\bibitem[An \& Evans(2016)]{An2016} An, J. \& Evans, N. W.\ 2016, \apj, 816, 35

\bibitem[Bensby, Feltzing \& Oey(2014)]{Bensby2014} Bensby, T., Feltzing, S., \& Oey, M. S.\ 2014, \aap, 562, A71

\bibitem[Binney \& Tremaine(1987)]{Binney1987} Binney, J., \& Tremain, S.\ 1987, Galactic Dynamics (Princeton, NJ: Princeton Univ. Press), 433

\bibitem[Binney \& Merrifield(1998)]{Binney1998} Binney, J., \& Merrifield, M.\ 1998, Galactic Astronomy (Princeton, NJ: Princeton Univ. Press)

\bibitem[Binney et al.(2014)]{Binney2014} Binney, J., Burnett, B., Kordopatis, G., et al.\ 2014, \mnras, 439, 1231

\bibitem[Bovy et al.(2016)]{Bovy2016} Bovy, J., Rix, H. W., Schlafly, E. F., et al.\ 2016, \apj, 823, 30

\bibitem[Brook et al.(2012)]{Brook2012} Brook, C. B., Stinson, G. S., Gibson, B. K., et al.\ 2012, \mnras, 426, 690

\bibitem[B{\"u}denbender et al.(2015)]{Budenbender2015} B{\"u}denbender, A., van de Ven, G., Watkins, L. L.\ 2015, \mnras, 452, 956

\bibitem[Chen et al.(2019)]{Chen2019} Chen, B. Q., Huang, Y., Yuan, H. B., et al.\ 2019, \mnras, 483, 4277

\bibitem[Chiappini et al.(2015)]{Chiappini2015} Chiappini, C., Anders,F., Rodrigues, T. S., et al.\ 2015, \aap, 576, L12

\bibitem[Collett, Dutta \& Evans(1997)]{Collett1997} Collett, J. L., Dutta, S. N., \& Evans, N. W.\ 1997, \mnras, 285, 49


\bibitem[Cui et al.(2012)]{Cui2012} Cui, X. Q., Zhao, Y. H., Chu, Y. Q., et al.\ 2012,\ RAA, 12, 1197

\bibitem[De Silva et al.(2015)]{De Silva2015} De Silva, G. M., Freeman, K. C., Bland-Hawthorn, J., et al.\ 2015, \mnras, 449, 2604

\bibitem[Dehnen(2000)]{Dehnen2000} Dehnen, W.\ 2000, \aj, 119, 800

\bibitem[Deng et al.(2012)]{Deng2012} Deng, L. C., Newberg, H. J., Liu, C., et al.\ 2012,\ RAA, 12, 735



\bibitem[Eddington(1915)]{Eddington1915} Eddington, A. S.\ 1915, \mnras, 76, 37

\bibitem[Eilers et al.(2019)]{Eilers2019} Eilers, A. C., Hogg, D. W., Rix, H. W., et al.\ 2019, \apj, 871, 120

\bibitem[Evans et al.(2016)]{Evans2016} Evans, N. W., Sanders, J. L., Williams, A. A., et al.\ 2016, \mnras, 456, 4506

\bibitem[Everall et al.(2019)]{Everall2019} Everall, A., Evans, N. W., Belokurov, V., and Sch{\"o}nrich, R.\ 2019, \mnras, 489, 910

\bibitem[Gaia Collaboration et al.(2022a)]{Gaia2022a} Gaia Collaboration, Creevey, O. L.,  Sarro, L. M.,  Lobel, A., et al.\ 2022, arXiv:2206.05870 

\bibitem[Gaia Collaboration et al.(2022b)]{Gaia2022b} Gaia Collaboration, Recio-Blanco, A., Kordopatis, G., et al.\ 2022, arXiv:2206.05534


\bibitem[G{\' o}mez et al.(2013)]{Gomez2013} G{\' o}mez, F. A., Minchev, I., O’Shea, B. W., Beers, T. C., Bullock, J. S., Purcell, C. W.\ 2013, \mnras, 429, 159

\bibitem[Guiglion et al.(2015)]{Guiglion2015} Guiglion, G., Recio-Blanco, A., de Laverny, P., et al.\ 2015, \aap, 583, A91

\bibitem[Hagen et al.(2019)]{Hagen2019} Hagen, J. H. J., Helmi, A., de Zeeuw, P. T.,  and Posti, L.\ 2019, \aap, 629, A70

\bibitem[Hayden et al.(2020)]{Hayden2020} Hayden, M. R., Bland-Hawthorn, J., Sharma, S., et al.\ 2020, \mnras, 493, 2952

\bibitem[Haywood et al.(2013)]{Haywood2013} Haywood, M., Di Matteo, P., Lehnert, M. D., et al.\ 2013, \aap, 560, A109

\bibitem[Huang et al.(2016)]{Huang2016} Huang, Y., Liu, X. W., Yuan, H. B., et al.\ 2016, \mnras, 463, 2623

\bibitem[Huang et al.(2018a)]{Huang2018a} Huang, Y., Liu, X.-W., Chen, B.-Q., et al.\ 2018a, \aj, 156, 90

\bibitem[Huang et al.(2018b)]{Huang2018b} Huang, Y., Sch{\"o}nrich, R., Liu, X. W., et al.,\ 2018b, \apj, 864, 129

\bibitem[Huang et al.(2020)]{Huang2020} Huang, Y., Sch{\"o}nrich, R., Zhang, H. W., et al.\ 2020, ApJS, 249, 29

\bibitem[Laporte et al.(2019)]{Laporte2019} Laporte, C. F. P., Minchev, I., Johnston, K. V., G{\' o}mez, F. A.\ 2019, \mnras, 485, 3134

\bibitem[Liu et al.(2014)]{Liu2014} Liu, X. W., Yuan, H. B., Huo, Z. Y., et al.\ 2014, in IAU Symp. 298, Setting the Scene for Gaia and LAMOST (Cambridge: Cambridge Univ. Press), 310

\bibitem[Majewski et al.(2017)]{Majewski2017} Majewski, S. R., Schiavon, R. P., Frinchaboy, P. M., et al.\ 2017, \aj, 154, 94

\bibitem[Minchev et al.(2015)]{Minchev2015} Minchev, I., Martig, M., Streich, D., et al.\ 2015, \apj, 804, L9

\bibitem[Momany et al.(2006)]{Momany2006} Momany, Y., Zaggia, S., Gilmore, G., et al.\ 2006, \aap, 451, 515

\bibitem[Binney \& Tremaine(2008)]{Binney2008} Binney, J., \& Tremaine, S.\ 2008, Galactic Dynamics (2nd ed.; Princeton, NJ: Princeton Univ. Press)

\bibitem[Bland-Hawthorn \& Gerhard(2016)]{Bland-Hawthorn2016} Bland-Hawthorn, J., Gerhard, O.\ 2016, \araa, 54, 529

\bibitem[Jofr{\'e} et al.(2016)]{Jofre2016} Jofr{\'e}, P., Jorissen, A., Van Eck, S., et al.\ 2016, \aap, 595, A60

\bibitem[Lee et al.(2011)]{Lee2011} Lee, Y. S., Beers, T. C., An, D., et al.\ 2011, \apj, 738, 187

\bibitem[Recio-Blanco et al.(2014)]{Recio-Blanco2014} Recio-Blanco, A., de Laverny, P., Kordopatis, G., et al.\ 2014, \aap, 567, A5

\bibitem[Smith et al.(2009)]{Smith2009} Smith, M. C., Evans, N. W., An, J. H.\ 2009, \apj, 698, 1110

\bibitem[Steinmetz et al.(2006)]{Steinmetz2006} Steinmetz, M., Zwitter, T., Siebert, A., et al.\ 2006, \aj, 132, 1645

\bibitem[Sun et al.(2020)]{Sun2020} Sun, W. X., Huang, Y., Wang, H. F., et al.\ 2020, \apj, 903, 12
\bibitem[Sun et al.(2023, in preparation)]{Sun2023} Sun, W. X., et al. 2013, in preparation

\bibitem[Yuan et al.(2015)]{Yuan2015} Yuan, H. B., Liu, X. W., Huo, Z. Y., et al.\ 2015, \mnras, 448, 855

\bibitem[Reid et al.(2014)]{Reid2014} Reid, M. J., Menten, K. M., Brunthaler, A., et al.\ 2014, \apj, 783, 130

\bibitem[Sch{\"o}nrich \& Dehnen(2018)]{Schonrich2018} Sch{\"o}nrich, R., \& Dehnen, W.\ 2018, \mnras, 478, 3809

\bibitem[Recio-Blanco et al.(2022)]{Recio2022} Recio-Blanco, A., de Laverny, P., Palicio, P.~A., et al.\ 2022, arXiv:2206.05541.

\bibitem[Reid \& Brunthaler(2004)]{Reid2004} Reid, M. J., \& Brunthaler, A.\ 2004, \apj, 616, 872

\bibitem[Sch{\"o}nrich et al.(2010)]{Schonrich2010} Sch{\"o}nrich, R., Binney, J., \& Dehnen, W.\ 2010, \mnras, 403, 1829

\bibitem[Sch{\"o}nrich(2012)]{Schonrich2012} Sch{\"o}nrich, R.\ 2012, \mnras, 427, 274

\bibitem[Saha, Pfenniger \& Taam(2013)]{Saha2013} Saha, K., Pfenniger, D., Taam, R. E.\ 2013, \apj, 764, 123

\bibitem[Trick et al.(2021)]{Trick2021} Trick, W. H., Fragkoudi, F., Hunt, J. A. S., et al.\ 2021, \mnras, 500, 2645


\bibitem[Huang et al.(2015)]{Huang2015} Huang, Y., Liu, X. W., Yuan, H. B., et al.\ 2015, \mnras, 449, 162

\bibitem[Huang et al.(2016)]{Huang2016} Huang, Y., Liu, X. W., Yuan, H. B., et al.\ 2016, \mnras, 463, 2623

\bibitem[Williams et al.(2013)]{Williams2013} Williams, M. E. K., Williams, M., Binney, J., et al.\ 2013, \mnras, 436, 101

\bibitem[Xiang et al.(2017)]{Xiang2017} Xiang, M. S., Liu, X. W., Yuan, H. B., et al.\ 2017, \mnras, 467, 1890

\bibitem[Xu et al.(2015)]{Xu2015} Xu, Y., Newberg, H. J., Carlin, J. L., et al.\ 2015, \apj, 801, 105




\end{thebibliography}
\end{document}